\documentclass[12pt,english,floatfix,showkeys,superscriptaddress,aps,prd,preprint]{revtex4}
\usepackage[latin1]{inputenc}
\usepackage[T1]{fontenc}
\usepackage{lmodern}
\usepackage{amsmath}
\usepackage{amssymb}
\usepackage{graphicx}
\usepackage{esint}
\usepackage{longtable}
\usepackage{dcolumn}
\usepackage{babel}
\usepackage{csquotes}
\usepackage{color}

\usepackage{hyperref}
\hypersetup{
    colorlinks,
    citecolor=blue,
    filecolor=green,
    linkcolor=purple,
    urlcolor=red,
}

\usepackage{slashed}

\usepackage{hyperref}
\hypersetup{colorlinks,breaklinks,
			citecolor=[rgb]{0,0.0,1.0},
            urlcolor=[rgb]{0.0,0.0,1.0},
            linkcolor=[rgb]{0,0.5,0.9}}

\begin{document}

\title{Exact solution for a traversable wormhole in a curvature-coupled antisymmetric background field}

\author{R. V. Maluf}
\email{r.v.maluf@fisica.ufc.br}
\affiliation{Universidade Federal do Cear\'a (UFC), Departamento de F\'isica,\\ Campus do Pici, Fortaleza - CE, C.P. 6030, 60455-760 - Brazil.}


\author{C. R Muniz}
\email{celio.muniz@uece.br}
\affiliation{Universidade Estadual do Cear\'a (UECE), Faculdade de Educa\c{c}\~{a}o, Ci\^{e}ncias e Letras de Iguatu\\ Av. D\'ario Rabelo, s/n, 63.502-253, Iguatu-CE, Brazil.}


\date{\today}

\begin{abstract}

In this work, we study a traversable wormhole sourced by an ideal matter fluid with an antisymmetric 2-tensor background field coupled to gravity in a scenario of spontaneously broken Lorentz symmetry. Contrary to employed in the literature, we use a nonminimal curvature-coupling term $B^{\mu\nu}B^{\kappa\lambda}R_{\mu\nu\kappa\lambda}$ which incorporates all three kinds of Lorentz-violating coefficient for the pure-gravity sector of the minimal standard-model extension. We find that the wormhole is non-asymptotically globally flat and determine the allowed parameters of the theory, showing that the matter fluid must be necessarily anisotropic. We also analyze the energy conditions, checking their validity range and comparing them with those predicted by general relativity.
\end{abstract}


\maketitle

\section{Introduction}

The possibility of Lorentz symmetry breaking at the Planck scale is supported by some fundamental theories, including strings \cite{Kostelecky:1988zi,Kostelecky:1991ak}, loop quantum gravity \cite{Gambini:1998it,Bojowald:2004bb}, noncommutative spacetimes \cite{Carroll:2001ws}, and Horava-Lifshitz gravity \cite{Horava:2009uw}. A theoretical framework that can describe the low energy effects resulting from such symmetry breaking is the Standard-Model Extension (SME) proposed by Colladay and Kosteleck\'{y} in the late 1990s \cite{Colladay:1996iz,Colladay:1998fq}. The SME incorporates CPT- and Lorentz symmetry violation terms in all the usual standard model sectors. It was later expanded to take gravity into account as an effective theory, capable of making predictions that can be tested or verified observationally within current technological limits \cite{Kostelecky:2003fs}. Since its construction, the SME has been subject to several theoretical studies and experimental tests that made it possible to raise tight constraints on Lorentz invariance in nature. An updated compilation of all upper bounds can be found in \cite{Kostelecky:2008ts}.

In this context, it is of deep interest to explore the consequences of Lorentz violation in the gravitational scenario involving compact objects like black holes \cite{Casana:2017jkc,Lessa:2019bgi,Ding:2020kfr,Maluf:2020kgf} and topologically non-trivial structures as wormholes \cite{Ovgun:2018xys,Lessa:2020imi} as well as in cosmology \cite{Bertolami:2005bh,Capelo:2015ipa,Maluf:2021lwh} . These objects originally were found as solutions for the field equations of general relativity, unraveling unexpected connections between two remote regions of the spacetime \cite{Einstein:1935tc,Misner:1957mt,Ellis:1973yv}. Wormholes can also occur in spacetimes with arbitrary dimensions and in several topologies (See \cite{Dias:2010uh}, and references therein). They usually do not satisfy the energy conditions of the general relativity, being necessary some exotic matter as a source, like the Casimir energy \cite{Morris:1988tu,Garattini:2019ivd,Jusufi:2020rpw,Alencar:2021ejd}. Nevertheless, classical and quantum-modified theories of gravity can change this feature, which has been widely discussed in the recent literature \cite{Richarte:2007zz,Richarte:2009zz,Cuyubamba:2018jdl,Fu:2018oaq,Alencar:2021enh}. Still concerning such objects, a class of modified theories as those embodying background fields coupled to the gravity, with spontaneously Lorentz symmetry breaking, were analyzed in Refs. \cite{Ovgun:2018xys,Lessa:2020imi}.

Thus, in this letter, we will study a traversable wormhole immersed in an antisymmetric 2-tensor background field coupled to gravity and sourced by a matter fluid in a scenario of spontaneously Lorentz symmetry breaking \cite{Altschul:2009ae}. Differently from that was employed recently in \cite{Lessa:2020imi}, we will use another coupling to the antisymmetric field with the curvature, which plays the role of a pseudo-electric one. We will show that the matter fluid can be ideal; however, it must be anisotropic. We will also determine the allowed parameters of the theory, analyze the energy conditions and verify where they are satisfied or violated, compared with that predicted by general relativity.

\section{The Einstein-bumblebee gravity with an antisymmetric tensor}

Here we present and discuss the main aspects of the model adopted in this work. The chosen action involves the Einstein-Hilbert term of general relativity with the presence of an antisymmetric 2-tensor $B_{\mu\nu}=-B_{\nu\mu}$, and it is given by \cite{Altschul:2009ae}
\begin{equation}
S =\int d^{4}x\sqrt{-g}\left[\frac{1}{2\kappa}R-\frac{1}{12}H_{\mu\nu\lambda}H^{\mu\nu\lambda}-V+\frac{\xi_{1}}{2\kappa}B^{\kappa\lambda}B^{\mu\nu}R_{\kappa\lambda\mu\nu}+\mathcal{L}_{M}\right],\label{eq:act1}
\end{equation}
where $\kappa=8\pi G_{N}$ such that $G_{N}$ is the Newtonian gravitational
constant, and the coupling constant $\xi_{1}$ (with mass dimension $[\xi_1]=M^{-2}$ in natural units) represents a nonderivative gravitational
coupling to $B_{\mu\nu}$ that is linear in the curvature. The field-strength tensor $H_{\mu\nu\lambda}$ associated with $B_{\mu\nu}$ is defined by
\begin{equation}
H_{\mu\nu\lambda}=\partial_{\mu}B_{\nu\lambda}+\partial_{\lambda}B_{\mu\nu}+\partial_{\nu}B_{\mu\lambda},\label{Hfield}
\end{equation}
with $H_{\mu\nu\lambda}$ being invariant under the gauge transformation
$B_{\mu\nu}\rightarrow B_{\mu\nu}+\partial_{\mu}\Lambda_{\nu}-\partial_{\nu}\Lambda_{\mu}$.
Also, the potential $V$ is responsible for triggering spontaneous
Lorentz violation inducing a nonzero vacuum value $\left\langle B_{\mu\nu}\right\rangle =b_{\mu\nu}$
, and $\mathcal{L}_{M}$ handles the ordinary matter content that
will be specified later.

Our motivation in choosing the theory defined by the action (\ref{eq:act1}) is twofold. First, the coupling involving the constant $\xi_{1}$ can generate all three Lorentz-violating coefficient fields of the SME, usually denoted as $u$, $s^{\mu\nu}$, and $t^{\kappa\lambda\mu\nu}$ \cite{Bailey:2006fd}. In particular, the $t$-coefficient can only be expressed in terms of the field $B_{\mu\nu}$ when contracted with the Riemann curvature tensor from the $\xi_{1}$ coupling \cite{Altschul:2009ae}. Second, the phenomenological role of the $t$-coefficient is an intriguing issue still little explored in the literature since previous investigations involving solutions of black holes \cite{Lessa:2019bgi} and wormholes \cite{Lessa:2020imi} did not take into account this type of coupling present in (\ref{eq:act1}). Hence, our main focus of this letter is to determine the existence of a wormhole solution modified by the $\xi_{1}$ coupling, responsible for producing a nonzero vacuum value for $t^{\kappa\lambda\mu\nu}$.

The equations of motion for gravity can be obtained from (\ref{eq:act1})
by varying with respect to $g_{\mu\nu}$ and keeping the other fields
fixed. Thus, we get
\begin{equation}
G^{\mu\nu} =\kappa(T_{M})^{\mu\nu}+\kappa(T_{B})^{\mu\nu} +(T_{\xi_{1}})^{\mu\nu}.\label{eq:metricfe}
\end{equation}
On the left side in (\ref{eq:metricfe}), we have the usual Einstein
tensor $G_{\mu\nu}=R_{\mu\nu}-\frac{1}{2}Rg_{\mu\nu}$, while on the
right side are represented the energy-momentum tensors due to the
matter content $(T_{M})_{\mu\nu}$ and the contributions originating
due to $B_{\mu\nu}$ field that appear from the kinetic and potential
terms $(T_{B})_{\mu\nu}$ and the nonminimal coupling $(T_{\xi_{1}})_{\mu\nu}$,
respectively.

For $B$-terms, we find explicitly
\begin{align}
(T_{B})^{\mu\nu} & =\frac{1}{2}H^{\alpha\beta\mu}H_{\ \alpha\beta}^{\nu}-\frac{1}{12}g^{\mu\nu}H^{\alpha\beta\gamma}H_{\alpha\beta\gamma}\nonumber \\
 & -g^{\mu\nu}V+4B^{\alpha\mu}B_{\alpha}^{\ \nu}V',
\end{align}
where for simplicity we have assumed the dependence of the potential
$V$ with respect to $B_{\mu\nu}$ through the form
\begin{equation}
V\equiv V(B_{\mu\nu}B^{\mu\nu}-x),
\end{equation}
with $x$ being a real number representing the vacuum value of the
invariant
\begin{equation}
x\equiv\left\langle B_{\mu\nu}B^{\mu\nu}\right\rangle =\left\langle g^{\alpha\mu}\right\rangle \left\langle g^{\beta\nu}\right\rangle b_{\alpha\beta}b_{\mu\nu},
\end{equation}
and the prime $'$ means derivative with respect to the potential
argument. Note that $\left\langle g^{\mu\nu}\right\rangle $ is the
vacuum value to the inverse metric. For our present purpose, we can
assume the $B_{\mu\nu}$ field and the metric are frozen in their
vacuum values so that
\begin{equation}
B_{\mu\nu}=b_{\mu\nu},\ \ \ \ g_{\mu\nu}=\left\langle g_{\mu\nu}\right\rangle ,
\end{equation}
and the vacuum conditions $V=V'=0$ are guaranteed.

Finally, the contributions due to nonminimal gravitational coupling is
\begin{align}
(T_{\xi_{1}})^{\mu\nu} & =\xi_{1}\left(\frac{1}{2}g^{\mu\nu}B^{\alpha\beta}B^{\gamma\delta}R_{\alpha\beta\gamma\delta}+\frac{3}{2}B^{\beta\gamma}B^{\alpha\mu}R_{\ \alpha\beta\gamma}^{\nu}\right.\nonumber \\
 & +\frac{3}{2}B^{\beta\gamma}B^{\alpha\nu}R_{\ \alpha\beta\gamma}^{\mu}+\nabla_{\alpha}\nabla_{\beta}B^{\alpha\mu}B^{\nu\beta}\nonumber \\
 & +\left.\nabla_{\alpha}\nabla_{\beta}B^{\alpha\nu}B^{\mu\beta}\right).
\end{align}

The equations of motion for the antisymmetric tensor field are also
obtained from the action (\ref{eq:act1}). By varying this action
concerning $B_{\mu\nu}$, and this time keeping the metric and matter
fields fixed, we have
\begin{equation}
\nabla_{\alpha}H^{\alpha\mu\nu}=4V'B^{\mu\nu}-\frac{2\xi_{1}}{\kappa}B_{\alpha\beta}R^{\alpha\beta\mu\nu}.\label{eq:KReom}
\end{equation} Note that here we are explicitly disregarding any type of coupling
between the matter fields and $B_{\mu\nu}$. Such a possibility
may imply changes in the conservation of the conventional matter currents
ant it is beyond our present scope.

%
%

\section{Static spherical wormhole solution with spontaneous Lorentz breaking}

The Morris-Thorne geometry which represents a static and spherically
symmetric wormhole solution is described by the line element \cite{Morris-Thorne,Morris:1988tu}
\begin{equation}\label{metric}
    ds^{2}=-e^{2\Phi(r)}dt^{2}+\left(1-\frac{\Omega(r)}{r}\right)^{-1}dr^{2}+r^{2}d\theta^{2}+r^{2} \sin^{2}{\theta}d\phi^{2},
\end{equation}
where $\Phi(r)$ is called the redshift function, admittedly everywhere finite to avoid event horizons or singularities, and $\Omega(r)$ is the shape function of the wormhole. Additional conditions on $\Phi(r)$ and $\Omega(r)$ are required for a transversable wormhole solution. One of them is the existence of the minimum radius: $\Omega(r_{0})=r_{0}$, where $r=r_{0}$ is the radius of the throat of the wormhole. Another important requirement is the flare-out condition at the  throat: $\Omega(r)<r$, while $\Omega'(r_{0})<1$. Also, the tidal gravitational forces must be very small, i.e., $\left|\Phi\right|\ll 1$. For the remainder of this work, we will neglect the tidal force, assuming $\Phi(r)=0$. With this simple choice, we can determine an exact wormhole solution, as we will see below.

For the matter-energy content, we adopt a perfect fluid such that the energy-momentum tensor for the matter has the $(T_{M})^{\mu}_{\ \ \nu}=\mbox{diag}(-\rho,p_{r},p_{\theta},p_{\phi})$. It is worth emphasizing that the perfect fluid is not assumed to be isotropic since the radial and lateral pressures are not a priori equal.

Now let us configure the Lorentz-violating field. Following the Refs. \cite{Lessa:2019bgi,Lessa:2020imi}, we will restrict ourselves to the pseudo-electric configuration in which the field $B_{\mu\nu}$ is frozen in its vacuum expectation value $b_{\mu\nu}$, whose explicit form is given by
\begin{equation}
b_{\mu\nu}=b_{10}=-b_{01}=\frac{a}{\sqrt{1-\frac{\Omega(r)}{r}}}.\label{bvev}
\end{equation} In this way, the background field $b_{\mu\nu}$ has a constant norm, $b_{\mu\nu}b^{\mu\nu}=-2a^2$, where $a$ is a real and positive parameter. This setup preserves the spherical and static spacetime symmetry. Moreover, according to relations (\ref{Hfield}) and (\ref{bvev}), the field strength $H_{\mu\nu\lambda}$ is identically null, and it can be explicitly verified that the equations of motion for $B_{\mu\nu}$ expressed in (\ref{eq:KReom}) are automatically satisfied.

Indeed, assuming the vacuum conditions, i.e., $V'=0=V$ and the vev (\ref{bvev}) for $B_{\mu\nu}$, the relevant components obtained from Eq. (\ref{eq:KReom}) are $-\frac{4\xi_{1}}{\kappa} b_{01}R^{01\mu\nu}$. On the other hand, for the metric (\ref{metric}) with the zero tidal condition ($\Phi=0$), it can be shown that the nonzero components of the Riemann curvature tensor are
\begin{equation}
R_{1212}=-\left(\frac{\Omega-r\Omega'}{2r-2\Omega}\right),\ \ R_{1313}=\sin^{2}\theta R_{1212},\ \ R_{2323}=r\sin^{2}\theta\Omega,
\end{equation}and also those obtained by the allowed permutations of indices. Thus, it is clear that the $b_{01}R^{01\mu\nu}$ term is identically null for the configuration adopted by us.

After these preliminary considerations, the extended Einstein equations (\ref{eq:metricfe}) result in the following nonvanishing components:
\begin{equation}
    \frac{\Omega'}{r^{2}}-\frac{\lambda}{r^{3}}\left(r\Omega'+\Omega-2r\right)-\kappa\rho=0,\label{rho}
\end{equation}
\begin{equation}
    \frac{\Omega}{r^{3}}+\kappa p_{r}=0,\label{pr}
\end{equation}
\begin{equation}
    \frac{1}{2}\left(\Omega'-\frac{\Omega}{r}\right)+\kappa r^{2}p_{\theta}=0,\label{ptheta}
\end{equation}
\begin{equation}
    \sin^{2}\theta\left[\frac{1}{2}\left(\Omega'-\frac{\Omega}{r}\right)+\kappa r^{2}p_{\phi}\right]=0,\label{pphi}
\end{equation}where we have defined the Lorentz-violating parameter $\lambda=2\xi_{1}a^{2}$. Note that the signature of the Lorentz violation only appears in the temporal part (\ref{rho}) of the field equations.

From Eq. (\ref{pr}), one can express the radial pressure in terms of the shape function as
\begin{equation}
p_{r}=-\frac{\Omega}{\kappa r^{3}}.\label{pr2}
\end{equation}

Assuming the equation of state: $p_{r}=\omega\rho$, where $\omega$ is a dimensionless real parameter, we insert the relation (\ref{pr2}) into Eq. (\ref{rho}) to get
\begin{equation}
(1-\lambda)r\Omega'+\left(\frac{1}{\omega}-\lambda\right)\Omega+2\lambda r=0.
\end{equation}

Solving the above equation, we obtain the solution for the shape function as follows
\begin{equation}
\Omega(r)=\frac{1}{1+(1-2\lambda)\omega}\left[-2\lambda\omega r+(1+\omega)r_{0}\left(\frac{r_{0}}{r}\right)^{\frac{1-\lambda\omega}{(1-\lambda)\omega}}\right],\label{omega}
\end{equation}and as before-mentioned $\Omega(r_{0})=r_{0}$.

Inserting Eq. (\ref{omega}) into Eq. (\ref{pr2}), we get the energy density
\begin{equation}
\rho(r)=\frac{1}{\omega\kappa\left(1+\omega-2\lambda\omega\right)r^{3}}\left[2\lambda\omega r-(1+\omega)r_{0}\left(\frac{r_{0}}{r}\right)^{\frac{1-\lambda\omega}{(1-\lambda)\omega}}\right].\label{solRho}
\end{equation}

It is easy to verify that the solutions (\ref{omega}) and (\ref{solRho}), together with the equation of state to $\rho$ and $p_{r}$, satisfy the temporal (\ref{rho}) and radial (\ref{pr}) components of the Einstein equations. However, these solutions do not verify the other components (\ref{ptheta}-\ref{pphi}) to the isotropic condition, i.e., $p_{r}=p_{\theta}=p_{\phi}$, except for the usual case when $\lambda=0$. Thus, it is impossible to determine a specific value for $\omega$ that preserves the isotropy of the perfect fluid and still allows the Lorentz breaking. For wormhole solutions in the context of bumblebee gravity models, this anisotropic condition is a new result and differs from those already obtained in the literature \cite{Ovgun:2018xys, Lessa:2020imi}.

Nevertheless, we can still assume $p_{\theta}=p_{\phi}$ and use Eqs. (\ref{ptheta}) and (\ref{omega}) to fix the lateral pressure, thus
\begin{equation}
p_{\theta}(r)=p_{\phi}(r)=\frac{(1+\omega)r_{0}}{2\omega\kappa(1-\lambda)r^{2}}\left(\frac{r_{0}}{r}\right)^{\frac{1-\lambda\omega}{(1-\lambda)\omega}}.
\end{equation} As expected, at the Lorentz invariant regime, i.e., $\lambda\rightarrow 0$, we recover the isotropic case for $\omega= -1/3$, such that $\Omega(r)=r^{3}/r_{0}^{2}$ and $p_{r}=p_{\theta}=p_{\phi}=-1/\kappa r_{0}^{2}$, which represents a spacetime of constant curvature \cite{Cataldo:2016dxq}.

\begin{figure}[t]
\begin{center}
\includegraphics[width=0.5\textwidth]{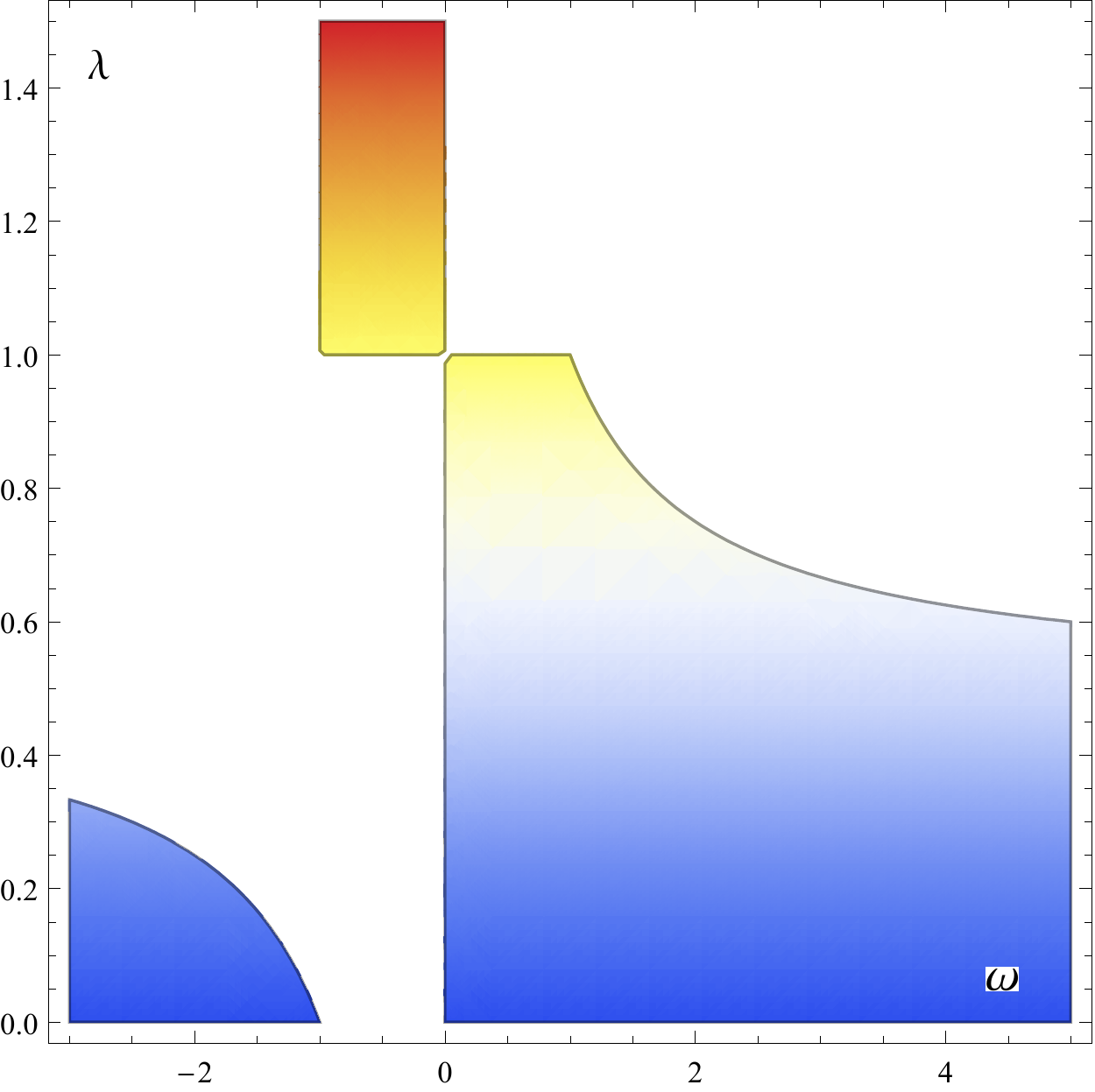}
\caption{Parameter space ($\omega$,$\lambda$) associated to the modified wormhole solution due to Lorentz breaking. The colored regions indicate the allowed values of these parameters for the formation of the wormhole. Hot (cold) colors indicate great (small) values for the Lorentz breakdown parameter, $\lambda$.}\label{fig1}
\end{center}
\end{figure}

One can easily see that the radial metric component $g_{rr}=\left(1-\frac{\Omega}{r}\right)^{-1}$ diverge at $r=r_{0}$, as is expected for any wormholes.  However,  our wormhole solution is non-asymptotically flat when $r\rightarrow\infty$:
\begin{equation}
\lim_{r\rightarrow\infty}\frac{\Omega(r)}{r}\Rightarrow-\frac{2\lambda\omega}{1+\omega-2\lambda\omega}+\frac{1+\omega}{1+\omega-2\lambda\omega}\lim_{r\rightarrow\infty}\left(\frac{r_{0}}{r}\right)^{1+\frac{1-\lambda\omega}{(1-\lambda)\omega}}.
\end{equation}
From the above result, it is clear that the first term does not depend on $r$, while the second term is vanished if
\begin{equation}
1+\frac{1-\lambda\omega}{(1-\lambda)\omega}>0,\label{omegacondition}
\end{equation}
otherwise the metric diverges at infinity. Generally speaking, the geometry of wormholes is asymptotically flat, at least in the general relativity context. Here we are dealing with a scenario of extended gravity, which comprises a classical background field which spontaneously violates the local Lorentz symmetry, and this explains the non-flatness at infinity. A similar result involving a wormhole solution in the presence of a vector background field was obtained in Ref. \cite{Ovgun:2018xys}. Such solutions have analogous asymptotic behavior to the spacetimes having topological defects. In fact, by a simple transformation of coordinates we can rewrite the wormhole metric at the asymptotic limit in the form,
\begin{equation}
ds^{2}=-dt^{2}+d\tilde{r}^{2}+\left(1+\frac{2\lambda\omega}{1+\omega-2\lambda\omega}\right)\tilde{r}^{2}(d\theta^{2}+\sin^{2}{\theta}d\phi^{2}),
\end{equation}where we take
\begin{equation}
\tilde{r}=\left(1+\frac{2\lambda\omega}{1+\omega-2\lambda\omega}\right)^{-1/2}r.
\end{equation}
The above result show that the asymptotic geometry corresponds to a global monopole \cite{Barriola:1989hx}, provided the condition (\ref{omegacondition}) is satisfied. Aligning this condition with the flare-out one, $\Omega(r)<r$ for all $r>r_0$, gives us the parameter space configuration shown in Figure \ref{fig1}, in which the allowed values for $\lambda$ and $\omega$ are exhibited, in thermal colors. Notice also that phantom-like fluids ($\omega<-1$), for little values of $\lambda$ can generate the wormhole under inspection.

From solution (\ref{omega}), we can calculate the Ricci and the Kretschmann scalars. Their values at $r=r_{0}$, are respectively,
\begin{equation}
\left.R\right|_{r=r_{0}}=-\frac{2(1+\lambda\omega)}{(1-\lambda)\omega}\frac{1}{r_{0}^{2}},\label{ricci}
\end{equation}and
\begin{equation}
\left.K\right|_{r=r_{0}}=\frac{2+4\omega+(6-8\lambda+4\lambda^{2})\omega^{2}}{(1-\lambda)^{2}\omega^{2}r_{0}^{4}}.\label{K}
\end{equation}
These quantities show that the wormholes are free of singularities at the throat, guaranteeing thus their traversability, for all $\lambda\neq 1$.

\begin{figure}[!ht]
    \centering
    \begin{minipage}{0.5\linewidth}
        \centering
        \includegraphics[width=1.1\textwidth]{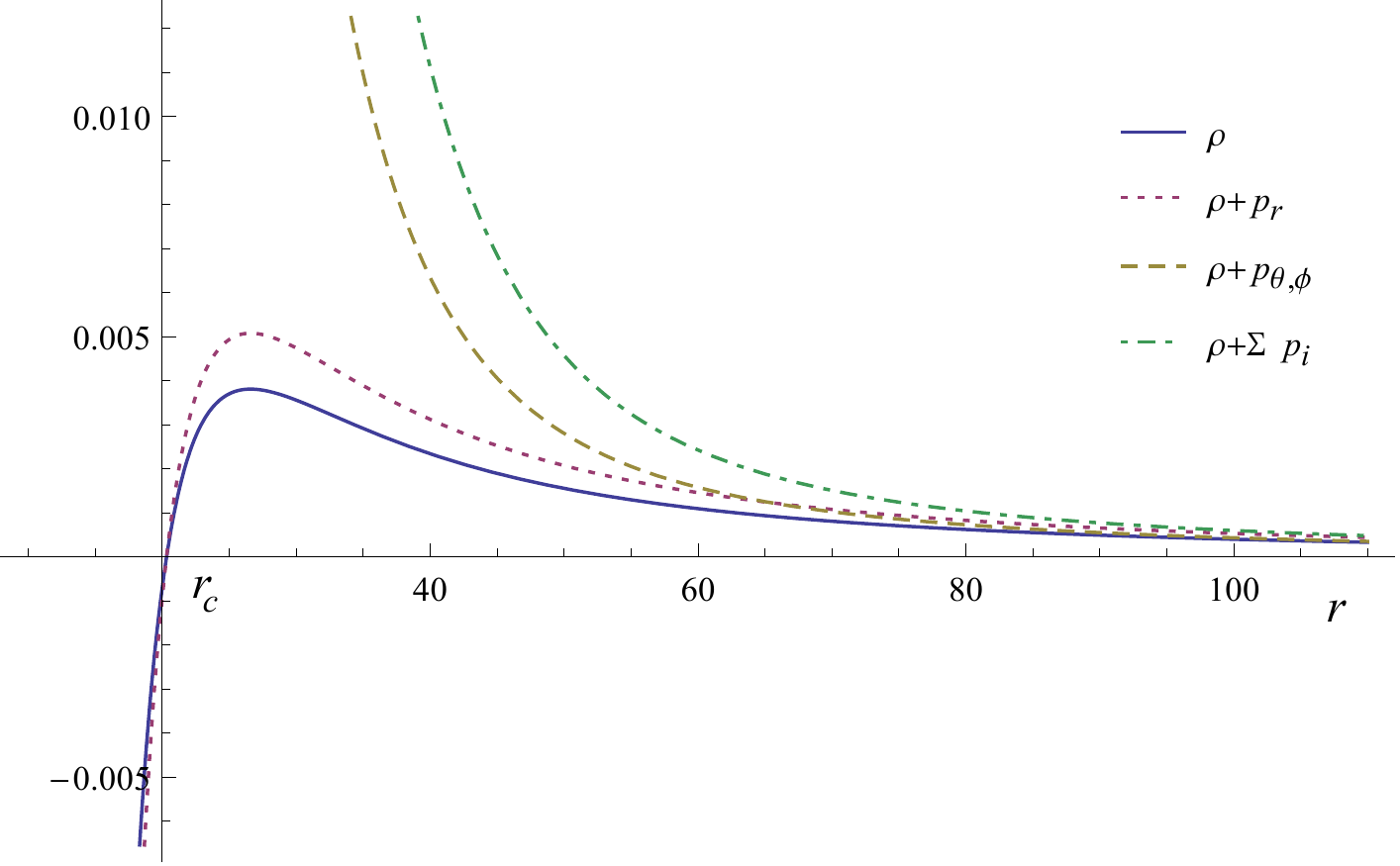}
        \label{fig:figura1minipg}
    \end{minipage}\hfill
    \begin{minipage}{0.5\linewidth}
        \centering
        \includegraphics[width=1.1\textwidth]{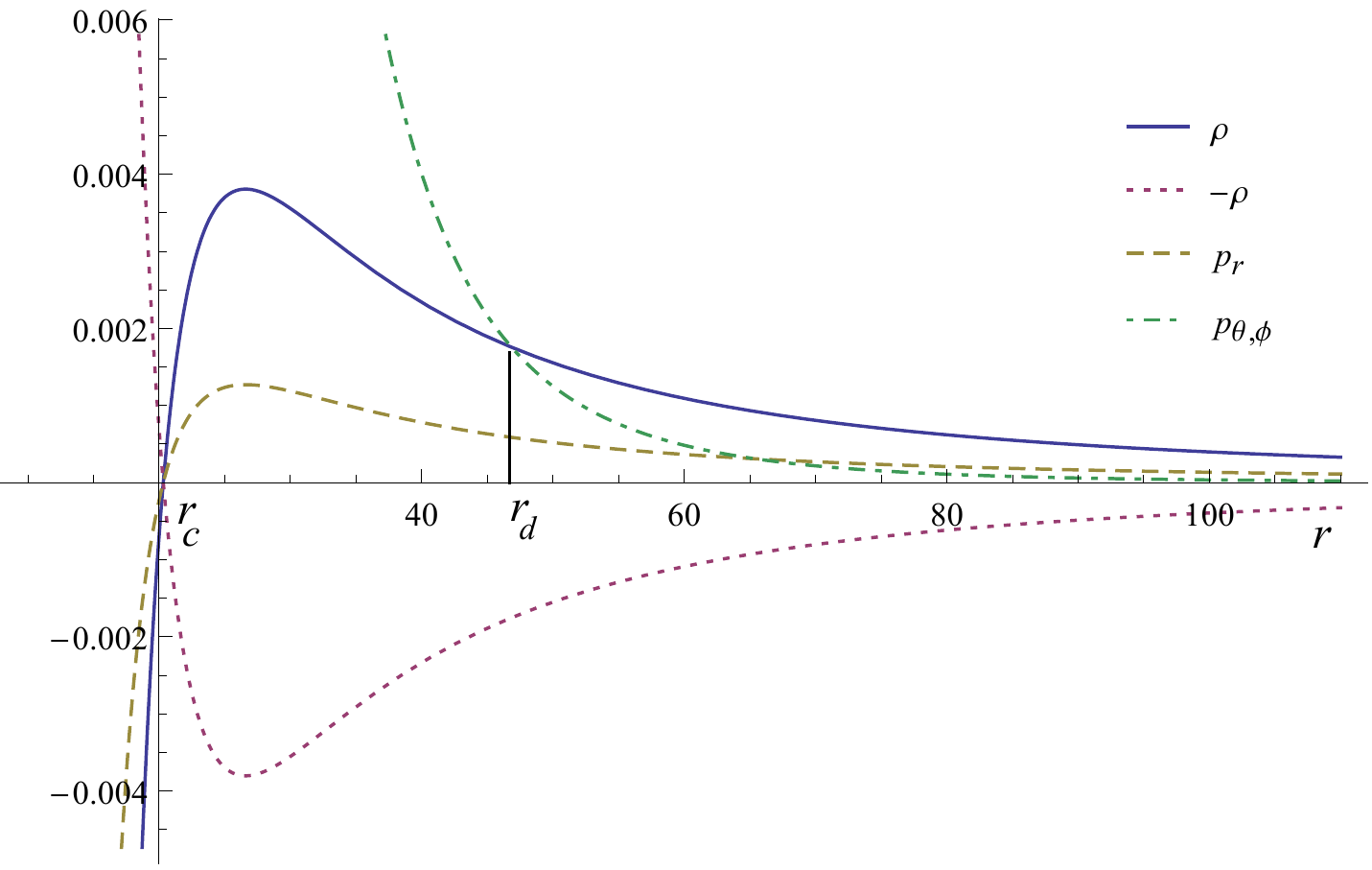}
              \label{fig:figura2minipg}
    \end{minipage}
   \caption{Energy density, pressures, and their combinations associated to the matter fluid, as functions of the radial coordinate, for $\lambda=0.1$, $\omega=1/3$, and $r_0=20$, in natural units.}
    \label{fig:figurasminipg}
\end{figure}

Lastly, we now analyze the energy conditions for the perfect anisotropic fluid supporting our modified wormhole solution. We depict in Figure \ref{fig:figurasminipg} some quantities taken from the energy density and pressures as functions of the radial coordinate, in order to verify the regions where the Null Energy Conditions (NEC, $\rho+p_i\geq 0$), Weak Energy Conditions (WEC, $\rho\geq 0$, $\rho+p_i\geq 0$), Strong Energy Conditions (SEC, $\rho+p_i\geq 0$, $\rho+\sum p_i\geq 0$), and Dominant Energy Conditions (DEC, $\rho\geq 0$, $-\rho\geq p_i\geq\rho$) are satisfied. We can notice that NEC, WEC, and SEC are entirely obeyed for \begin{equation}
r\geq r_c =  \left(\frac{2\omega \lambda}{\omega+1}\right)^{\frac{\omega (1-\lambda)}{2 \omega \lambda-w-1}}r_0, \label{Critical1}
\end{equation}
which is confirmed in the left panel of Figure \ref{fig:figurasminipg}.

On the other hand, those conditions as well as DEC are fully satisfied for $r\geq r_d$, which cannot be analytically determined. Notice however that $r_d > r_c$, and this can be clearly seen in the right panel of Figure \ref{fig:figurasminipg}. Thus, we conclude that only in the neighborhood of the wormhole throat all the energy conditions are violated. We also notice that on returning to general relativity by making $\lambda=0$, they leave be satisfied at all $r>r_0$. The present analysis shows us that even an ordinary fluid, like electromagnetic radiation ($\omega=1/3$), can contribute to the formation of the wormhole due to the presence of a background field that violates the Lorentz symmetry, behaving effectively like exotic matter nearby the wormhole throat.

\section{Conclusion\label{conclusion}}

In this work, we have studied traversable wormhole solutions with an antisymmetric 2-tensor background field coupled to gravity, with spontaneous Lorentz symmetry breaking. We have used a gravity coupling to that field still not employed in the literature, which by its turn plays the role of a pseudo-electric field, finding a traversable wormhole solution non-asymptotically globally flat. To ensure that all components of the modified Einstein equations are satisfied, we relax the isotropy assumption for the matter content and assume that the equation of state $p=\omega \rho$ is valid only for radial pressure. The solution obtained is then used to fix the lateral pressures.
 Also, we have established a parameter space where this must occur, together with the flare-out condition, for a perfect fluid sustaining the wormhole. Also, we have identified that even a phantom-like fluid can be a source for this modified wormhole solution. Furthermore, the matter fluid needs to be anisotropic, compatible with the nonvanishing vev of the background field itself. Such a feature was not found in the case examined in Ref. \cite{Lessa:2020imi}. Furthermore, we have found that all the energy conditions are satisfied beyond a critical radius. That is different from what happens in general relativity, in which those conditions leave to be valid everywhere. We have shown that such a violation occurs only near the wormhole throat and verifies that an ordinary fluid, as electromagnetic radiation, can be a source to this structure, behaving, thus, as an exotic fluid, due to the presence of the antisymmetric background field.
As a future perspective, we intend to study the influence of the pseudo-magnetic sector of the antisymmetric background field, together with the matter fluid in building a traversable wormhole.

\section*{Acknowledgments}
\hspace{0.5cm} The authors thank the Funda\c{c}\~{a}o Cearense de Apoio ao Desenvolvimento
Cient\'{i}fico e Tecnol\'{o}gico (FUNCAP), the Coordena\c{c}\~{a}o de Aperfei\c{c}oamento de Pessoal de N\'{i}vel Superior (CAPES), and the Conselho Nacional de Desenvolvimento Cient\'{i}fico e Tecnol\'{o}gico (CNPq), Grants no
307556/2018-2 (RVM) and PRONEM PNE-0112-00085.01.00/16 (CRM) for financial support.


\end{document}